# Maintenance Problem of Insufficiently Financed Pension Funds-A Stochastic Approach

*Manuel Alberto M. Ferreira*
Instituto Universitário de Lisboa (ISCTE-IUL),
Business Research Unit (BRU-IUL) and Information Sciences,
Technologies and Architecture Research Center (ISTAR-IUL),
Lisboa, Portugal

## Abstract

The generic case of pensions fund that it is not sufficiently auto financed and it is thoroughly maintained with an external financing effort is considered in this chapter. To represent the unrestricted reserves value process of this kind of funds, a time homogeneous diffusion stochastic process with finite expected time to ruin is proposed. Then it is projected a financial tool that regenerates the diffusion at some level with positive value every time the diffusion hits a barrier placed at the origin. So, the financing effort can be modeled as a renewal-reward process if the regeneration level is preserved constant. The perpetual maintenance cost expected values and the finite time maintenance cost evaluations are studied. An application of this approach when the unrestricted reserves value process behaves as a generalized Brownian motion process is presented.

## INTRODUCTION

The sustainability of a pensions fund study is performed, for instance, in Ferreira (2016), (Ferreira, Andrade, and Filipe, 2012) and (Figueira, and Ferreira, 1999) in the field of queuing theory. Two infinite servers' queues are considered: one with the contributors to the fund, which service time is the time during which they contribute to the fund; the other with the pensioners which service time is the time during which they receive the pension. In both queues, there is no distinction between a customer and its server in technical sense.

The category of fund portrayed herein is the so-called distributive pensions fund. Public pensions funds are generally of this kind. Their income is the contributions of the workers that are distributed by pensioners in the form of retirement pensions. The most important result of these studies is that, for the fund to be in balance, the average pension should be equal to the average contribution. However, due to the demographic imbalance that is taking shape in contemporary societies, with the number of taxpayers decreasing successively while the number of pensioners increases inexorably, it has meant that if the funds were to be autonomously balanced the contributions of the workers would have to assume unbearable values.

What usually happens is the injection of capital into these funds through transfers from the Public Budget whenever necessary. Therefore, in a disorganized way and, in general, in unforeseen situations that may coincide with large financial difficulties. The purpose of this study is to try to make these situations more predictable, both in relation to the time of occurrence and the amount needed, so that the protection of these funds occurs as smoothly as possible.

Along this chapter, this problem is approached bearing in mind the protection cost present value expectation for a non-autonomous pensions fund. Two problems are considered in this context:

- The above-mentioned expectation when the protection effort is perpetual,
- The protection effort for a finite time.

It is admitted that the unrestricted fund reserves behavior may be modeled as a time homogeneous diffusion process and use then a regeneration scheme of the diffusion to include the effect of an external financing effort.

A comparable work may be seen in (Gerber, and Parfumi, 1998), where has been considered a Brownian motion process conditioned by a reflection scheme. Less constrained, but in different conditions, exact solutions were then obtained for both problems.

The work presented in Refait (2000), on asset-liability management aspects, motivated the use of an application of the Brownian motion example in that domain.

Part of this work was presented at the Fifth International Congress on Insurance: Mathematics & Economics (Figueira, and Ferreira, 2001). Other works on this subject are (Figueira, and Ferreira, 2000, 2003), Ferreira (2012) and (Ferreira, and Filipe, 2013).

## REPRESENTATION OF THE PENSIONS FUNDS RESERVES BEHAVIOUR

Be $X(t), t \geq 0$ the reserves value process of a pensions fund given by an initial reserve amount $a, a > 0$ added to the difference between the total amount of contributions received up to time $t$ and the total amount paid in pensions up to time $t$. It is assumed that $X(t)$ is a time homogeneous diffusion process, with $X(0) = a$, defined by drift and diffusion coefficients: $\lim_{h \to 0} \frac{1}{h} E[X(t+h) - X(t)|X(t) = x] = \mu(x)$ and $\lim_{h \to 0} \frac{1}{h} E\left[(X(t+h) - X(t))^2 |X(t) = x\right] = \sigma^2(x)$.

Call $S_a$ the first passage time of $X(t)$ by 0, coming from $a$. The funds to be considered in this work are non-autonomous funds. So

$$E[S_a] < \infty, \text{ for any } a > 0 \tag{2.1}$$

that is: funds where the pensions paid consume in finite expected time any initial positive reserve and the contributions received, so that other financing resources are needed in order that the fund survives.

The condition (2.1) may be fulfilled for a specific diffusion process using criteria based on the drift and diffusion coefficients. Here the work presented in (Bhattacharya, and Waymire, 1990), pg. 418-422, is followed in this context. Begin accepting that $P(S_a < \infty) = 1$ if the diffusion scale function $q(x) = \int_{x_0}^{x} e^{-\int_{x_0}^{z} \frac{2\mu(y)}{\sigma^2(y)} dy} dz$, where $x_0$ is a diffusion state space fixed arbitrary point, fulfilling $q(\infty) = \infty$. Then the condition (2.1) is equivalent to $p(\infty) < \infty$, where $p(x) = \int_{x_0}^{x} \frac{2}{\sigma^2(z)} e^{\int_{x_0}^{z} \frac{2\mu(y)}{\sigma^2(y)} dy} dz$, is the diffusion speed function.

It is admitted that whenever the exhaustion of the reserves happens an external source places instantaneously an amount $\theta, \theta > 0$ of money in the fund so that it may keep on performing its function.

The reserves value process conditioned by this financing scheme is represented by the modification $\check{X}(t)$ of $X(t)$ that restarts at the level $\theta$ whenever it hits 0. Note that since $X(t)$ was defined as a time homogeneous diffusion, $\check{X}(t)$ is a regenerative process. Call $T_1, T_2, T_3, \ldots$ the sequence of random variables where $T_n$ denotes the $n^{th}$ $\check{X}(t)$ passage time by 0. It is obvious that the sequence of time intervals between these hitting times $D_1 = T_1, D_2 = T_2 - T_1, D_3 = T_3 - T_2, \ldots$ is a sequence of independent random variables where $D_1$ has the same probability distribution as $S_a$ and $D_2, D_3, \ldots$ the same probability distribution as $S_\theta$.

## FIRST PASSAGE TIMES LAPLACE TRANSFORMS

Call $f_a(s)$ the probability density function of $S_a(D_1)$. The corresponding probability distribution function is denoted by $F_a(s)$. The Laplace transform of $S_a$ is $\varphi_a(\lambda) = E[e^{-\lambda S_a}] = \int_0^\infty e^{-\lambda s} f_a(s) ds, \lambda > 0$.

Consequently, the density, distribution and transform of $S_\theta$ $(D_2, D_3, \ldots)$ will be denoted by $f_\theta(s), F_\theta(s)$ and $\varphi_\theta(\lambda)$, respectively.

The transform $\varphi_a(\lambda)$ satisfies the second order differential equation

$$\frac{1}{2}\sigma^2(a)u_\lambda''(a) + \mu(a)u_\lambda'(a) = \lambda u_\lambda(a), u_\lambda(a) = \varphi_a(\lambda), u_\lambda(0) = 1,$$
$$u_\lambda(\infty) = 0$$

(3.1)

see (Feller, 1971), pg. 478, (Karlin, and Taylor, 1981), pg. 243, and (Bass, 1998), pg. 89.

## PERPETUAL MAINTENANCE COST PRESENT VALUE

Consider the perpetual maintenance cost present value of the pensions fund that is given by the random variable $V(r, a, \theta) =$

$\sum_{n=1}^{\infty} \theta e^{-rT_n}$, r>0, where $r$ represents the appropriate discount rate. Note that $V(r, a, \theta)$ is a random perpetuity: what matters is its expected value which is easy to get using Laplace transforms. Since the $T_n$ Laplace transform is $E[e^{-\lambda T_n}] = \varphi_a(\lambda)\varphi_\theta^{n-1}(\lambda)$,

$$v_r(a, \theta) = E[V(r, a, \theta)] = \frac{\theta \varphi_a(r)}{1 - \varphi_\theta(r)} \tag{4.1}$$

It is relevant to note[1] that

$$\lim_{\theta \to 0} v_r(a, \theta) = \frac{u_r(a)}{-u_r'(0)}. \tag{4.2}$$

## FINITE TIME MAINTENANCE COST PRESENT VALUE

Define the renewal process $N(t)$, generated by the extended sequence $T_0 = 0, T_1, T_2, \ldots$, by $N(t) = \sup\{n: T_n \leq t\}$. The present value of the pensions fund maintenance cost up to time $t$ is represented by the stochastic process $W(t; r, a, \theta) = \sum_{n=1}^{N(t)} \theta e^{-rT_n}, W(t; r, a, \theta) = 0 \ if \ N(t) = 0$.

The important now is the expected value function of the process evaluation: $w_r(t; a, \theta) = E[W(t; r, a, \theta)]$. Begin to note that it may be expressed as a numerical series. Indeed, evaluating the expected value function conditioned by $N(t) = n$, it is obtained $E[W(t; r, a, \theta)|N(t) = n] = \theta \varphi_a(r) \frac{1 - \varphi_\theta^n(r)}{1 - \varphi_\theta(r)}$.

Repeating the expectation:

$$w_r(t; a, \theta) = E[E[W(t; r, a, \theta)]|N(t)] = \theta \varphi_a(r) \frac{1 - \gamma(t, \varphi_\theta(r))}{1 - \varphi_\theta(r)}$$

---

[1] Using the alternative notation, that seems more convenient now.

$$(5.1)$$

where $\gamma(t,\xi)$ is the probability generating function of $N(t)$.

Denote now the $T_n$ probability distribution function by $G_n(s)$ and assume $G_0(s) = 1$, for $s \geq 0$. Recalling that $P(N(t) = n) = G_n(t) - G_{n+1}(t)$, the above-mentioned probability generating function is

$$\gamma(t,\xi) = \sum_{n=0}^{\infty} \xi^n$$
$$P(N(t) = n) = 1 - (1-\xi) \sum_{n=1}^{\infty} \xi^{n-1} G_n(t) \quad (5.2).$$

Entering with (5.2) in (5.1), $w_r(t; a, \theta)$ is expressed in the form of the series

$$w_r(t; a, \theta) = \theta \varphi_a(r) \sum_{n=1}^{\infty} \varphi_\theta^{n-1}(r) G_n(t) \quad (5.3)$$

Then, using (5.3), it will be stated that $w_r(t; a, \theta)$ satisfies a renewal type integral equation.

Write for the $w_r(t; a, \theta)$ ordinary Laplace transform $\psi(\lambda) = \int_0^\infty e^{-\lambda s} w_r(s; a, \theta) ds$. Recalling that the probability distribution function $G_n(s)$ of $T_n$ ordinary Laplace transform is given by $\int_0^\infty e^{-\lambda s} G_n(s) ds = \varphi_a(\lambda) \frac{\varphi_\theta^{n-1}(\lambda)}{\lambda}$ and performing the Laplace transforms in both sides of (5.3) it is obtained $(\lambda) = \frac{\theta \varphi_a(r) \varphi_a(\lambda)}{\lambda(1 - \varphi_\theta(r) \varphi_\theta(\lambda))}$, that is

$$\psi(\lambda) = \theta \varphi_a(r) \frac{\varphi_a(\lambda)}{\lambda} + \psi(\lambda) \varphi_\theta(r) \varphi_\theta(\lambda) \quad (5.4)$$

Inverting the transforms in this equation both sides of (5.4) the following defective renewal equation

$$w_r(t; a, \theta) = \theta \varphi_a(r) F_a(t) + \int_0^t w_r(t-s; a, \theta) \varphi_\theta(r) f_\theta(s) ds$$
$$(5.5.)$$

is achieved.

Now an asymptotic approximation of $w_r(t; a, \theta)$ will be obtained through the key renewal theorem, following (Feller, 1971), pg. 376.

If in (5.5) $t \to \infty$

$$w_r(\infty; a, \theta) = \theta \varphi_a(r) + w_r(\infty; a, \theta) \varphi_\theta(r) \tag{5.6}$$

or $w_r(\infty; a, \theta) = \frac{\theta \varphi_a(r)}{1 - \varphi_\theta(r)} = v_r(a, \theta)$.

That is: the expression (4.1) for $v_r(a, \theta)$ is obtained again. Subtracting each side of (5.6) from the corresponding each side of (5.5), and performing some elementary calculations the following, still defective, renewal equation

$$J(t) = j(t) + \int_0^t J(t-s) \varphi_\theta(r) f_\theta(s) ds \tag{5.7}$$

where $J(t) = w_r(\infty; a, \theta) - w_r(t; a, \theta)$ and $j(t) = \theta \varphi_a(r)\big(1 - F_a(t)\big) + \frac{\theta \varphi_a(r) \varphi_\theta(r)}{1 - \varphi_\theta(r)} \big(1 - F_\theta(t)\big)$ results.

Now, to obtain a common renewal equation from (5.7), it must be admitted the existence of a value $k > 0$ such that $\int_0^\infty e^{ks} \varphi_\theta(r) f_\theta(s) ds = \varphi_\theta(r) \varphi_\theta(-k) = 1$.

This imposes that the transform $\varphi_\theta(\lambda)$ is defined in a domain different from the one initially considered, that is a domain that includes a convenient subset of the negative real numbers.

Multiplying both sides of (5.7) by $e^{kt}$ the common renewal equation desired is finally obtained: $e^{kt} J(t) = e^{kt} j(t) + \int_0^t e^{k(t-s)} J(t-s) e^{ks} \varphi_\theta(r) f_\theta(s) ds$ from which, through the application of the key renewal theorem, it results

$$\lim_{t \to \infty} e^{kt} J(t) = \frac{1}{k_0} \int_0^\infty e^{ks} j(s) ds \tag{5.8}$$

with $k_0 = \int_0^\infty s e^{ks} \varphi_\theta(r) f_\theta(s) ds = \varphi_\theta(r) \varphi_\theta'(-k)$, provided that $e^{kt} j(t)$ is directly Riemann integrable. The integral in (5.8) may expressed in terms of transforms as $\int_0^\infty e^{ks} j(s) ds = \frac{\theta \varphi_a(r) \varphi_a(-k)}{k}$.

So, in this section:

- An asymptotic approximation, in the sense of (5.8) was obtained:

$$w_r(t; a, \theta) \approx v_r(a, \theta) - c_r(a, \theta) e^{-k_r(\theta) t} \qquad (5.9)$$

where $k_r(\theta)$ is the positive value of $k$, solution of the following equation

$$\varphi_\theta(r) \varphi_\theta(-k) = 1 \qquad (5.10)$$

and

$$c_r(a, \theta) = \frac{\theta \varphi_a(r) \varphi_a(-k_r(\theta))}{-k_r(\theta) \varphi_\theta(r) \varphi_\theta'(-k_r(\theta))} \qquad (5.11)$$

## GENERALIZED BROWNIAN MOTION APPROACH

Consider the diffusion process $X(t)$ underlying the reserves value behavior of the pensions fund is a generalized Brownian motion process, with drift $\mu(x) = \mu, \mu < 0$ and diffusion coefficient $\sigma^2(x) = \sigma^2, \sigma > 0$. Observe that the setting satisfies the conditions that were assumed before to the former work, namely $\mu < 0$ implies condition (2.1). Everything else remaining as previously stated, it will be proceeded to present the consequences of this particularization. In general, it will be added a ∗ to the notation used before because it is intended to use these specific results later.

To get the first passage time $S_a$ Laplace transform it must be solved (3.1). This is a homogeneous second order differential equation with constant coefficients, which general solution is given by $u_\lambda^*(a) = \beta_1 e^{\alpha_1 a} + \beta_2 e^{\alpha_2 a}$, with $\alpha_1, \alpha_2 = \frac{-\mu \pm \sqrt{\mu^2 + 2\lambda \sigma^2}}{\sigma^2}$. Condition $u_\lambda^*(\infty) = 0$ implies $\beta_1 = 0$ and $u_\lambda^*(0)=1$ implies $\beta_2=1$ so that the following particular solution is achieved:

$$u_\lambda^*(a) = e^{-K_\lambda a} \left(= \varphi_a^*(\lambda)\right), K_\lambda = \frac{\mu + \sqrt{\mu^2 + 2\lambda \sigma^2}}{\sigma^2} \tag{6.1}$$

For this situation, the perpetual maintenance cost present value of the pensions fund is given by, following (4.1) and using (6.1),

$$v_r^*(a, \theta) = \frac{\theta e^{-K_r a}}{1 - e^{-K_r \theta}} \tag{6.2}$$

Note that $v_r^*(a, \theta)$ is a decreasing function of $a$ and an increasing function of $\theta$. Proceeding as before, in particular

$$\lim_{\theta \to 0} v_r^*(a, \theta) = \frac{e^{-K_r a}}{K_r}. \tag{6.3}$$

To reach an expression for the finite time period maintenance cost present value, start by the computation of $k_r^*(\theta)$, solving (5.10). This means finding a positive $k$ satisfying $e^{-K_r \theta} e^{-K_{-\lambda} \theta} = 1$ or $K_r + K_{-\lambda} = 0$. This identity is verified for the value of $k$

$$k_r^*(\theta) = \frac{\mu^2 - \left(-2\mu - \sqrt{\mu^2 + 2r\sigma^2}\right)^2}{2\sigma^2}, \text{if } \mu < -\sqrt{\frac{2r\sigma^2}{3}} \tag{6.4}$$

Note that the solution is independent of $\theta$ in these circumstances. A simplified solution, independent of $a$ and , for $c_r^*(a,\theta)$ was also obtained. Using (5.11) the result is

$$c_r^*(a,\theta) = \frac{2\sigma^2\left(-2\mu-\sqrt{\mu^2+2r\sigma^2}\right)}{\mu^2-\left(-2\mu-\sqrt{\mu^2+2r\sigma^2}\right)^2} \qquad (6.5)$$

Combining these results as in (5.9) it is observable that the asymptotic approximation for this particularization reduces to $w_r^*(t;a,\theta) \approx v_r^*(a,\theta) - \pi_r(t)$, where the function $\pi_r(t)$ is, considering (6.4) and (6.5),

$$\pi_r(t) = \frac{2\sigma^2\left(-2\mu-\sqrt{\mu^2+2r\sigma^2}\right)}{\mu^2-\left(-2\mu-\sqrt{\mu^2+2r\sigma^2}\right)^2} \, e^{-\frac{\mu^2-\left(-2\mu-\sqrt{\mu^2+2r\sigma^2}\right)^2}{2\sigma^2}t},$$
$$\text{if } \mu < -\sqrt{\frac{2r\sigma^2}{3}} \qquad (6.6)$$

## ASSETS AND LIABILITY BEHAVIOUR REPRESENTATION

It is proposed to consider now an application of the results obtained earlier to an asset-liability management scheme of pensions fund. Assume that the assets value process of a pensions fund may be represented by the geometric Brownian motion process $A(t) = be^{a+(\rho+\mu)t+\sigma B(t)}$ with $\mu < 0$ and $ab\rho + \mu\sigma > 0$, where $B(t)$ a standard Brownian motion process is. Suppose also that the liabilities value process of the fund performs as the deterministic process $L(t) = be^{\rho t}$.

Under these assumptions, consider now the stochastic process $Y(t)$ obtained by the elementary transformation of $A(t)$: $Y(t) = \ln\frac{A(t)}{L(t)} = a + \mu t + \sigma B(t)$.

This is a generalized Brownian motion process exactly as the one studied before, starting at $a$ and with drift $\mu$ and diffusion coefficient $\sigma^2$. Note also that the first passage time of the assets process $A(t)$ by the mobile barrier $T_n$, the liabilities process, is the first passage time of $Y(t)$ by 0-with finite expected time under the condition, stated before, $\mu < 0$.

Consider also the pensions fund management scheme that raises the assets value by some positive constant $\theta_n$ when the assets value falls equal to the liabilities process by the $n^{th}$ time. This corresponds to consider the modification $\bar{A}(t)$ of the process $A(t)$ that restarts at times $T_n$ when $A(t)$ hits the barrier $L(t)$ by the $n^{th}$ time at the level $L(T_n) + \theta_n$. For purposes of later computations, it is a convenient choice the management policy where

$$\theta_n = L(T_n)(e^\theta - 1), \text{ for some } \theta > 0 \tag{7.1}$$

The corresponding modification $\tilde{Y}(t)$ of $Y(t)$ will behave as a generalized Brownian motion process that restarts at the level $ln \frac{L(T_n)+\theta_n}{L(T_n)} = \theta$ when it hits 0 (at times $T_n$).

Proceeding this way, it is reproduced via $\tilde{Y}(t)$ the situation observed before when the Brownian motion example was treated. The Laplace transform in (6.1) is still valid.

As to former proceedings, the results for the present case will be denoted with the symbol #. It is considered the pensions fund perpetual maintenance cost present value, as a consequence of the proposed asset-liability management scheme, given by the random variable: $V^\#(r, a, \theta) = \sum_{n=1}^{\infty} \theta_n e^{-rT_n} = \sum_{n=1}^{\infty} b(e^\theta - 1)e^{-(r-\rho)T_n}, r > \rho$ where $r$ represents the appropriate discount interest rate. To obtain the above expression it was only made use of the $L(t)$ definition and (7.1). It is possible to express the expected value of the above random variable with the help of (6.2) as

$$v_r^\#(a, \theta) = \frac{b(e^\theta-1)}{\theta} v_{r-\rho}^*(a, \theta) = \frac{b(e^\theta-1)e^{-K_{r-\rho}a}}{1-e^{-K_{r-\rho}\theta}} \tag{7.2}$$

As $\theta \to 0$

$$\lim_{\theta \to 0} v_r^\#(a, \theta) = \frac{be^{-K_{r-\rho}a}}{K_{r-\rho}} \tag{7.3}$$

In a similar way, the maintenance cost up to time *t* in the above-mentioned management scheme, is the stochastic process

$$W^\#(t; r, a, \theta) = \sum_{n=1}^{N(t)} b(e^\theta - 1)e^{-(r-\rho)T_n}, \quad W^\#(t; r, a, \theta) = 0 \text{ if } N(t) = 0,$$ with expected value function

$$w_r^\#(t; a, \theta) = \frac{b(e^\theta - 1)}{\theta} w_{r-\rho}^*(t; a, \theta) \tag{7.4}$$

## CONCLUSION

In the general diffusion scenery, the main results are formulae (4.1) and (5.9). The whole work depends on the possibility of solving equation (3.1) to obtain the first passage times Laplacce transforms. Unfortunately, the solutions are known only for rare cases. An obvious case for which the solution of the equation is available is the one of the Brownian motion diffusion process. The main results concerning this particularization are formulae (6.2) and (6.6). Some transformations of the Brownian motion process that allowed to make use of the available Laplace transform may be explored as it was done in section 7. Formulae (7.2) and (7.4) are this application most relevant results.